\documentclass[10pt,a4paper]{article}
\usepackage[left=18mm, top=10mm, right=14mm, bottom=20mm, nohead, foot=10mm]{geometry}
\usepackage[affil-it]{authblk}

\let\OLDthebibliography\thebibliography
\renewcommand\thebibliography[1]{
  \OLDthebibliography{#1}
  \small
  \setlength{\parskip}{0pt}
  \setlength{\itemsep}{0pt plus 0.3ex}
}

\usepackage{multicol}

\usepackage{caption}

\usepackage{misccorr}

\usepackage[english]{babel}
\usepackage{amsmath,amssymb}
\usepackage{textcomp}
\usepackage{ulem}

\usepackage{xcolor}

\usepackage{hyperref}

\hypersetup{colorlinks=true, linkcolor=[rgb]{0.6 0 0}, citecolor=[rgb]{0 0.4 0}, urlcolor=[rgb]{0 0 0.6}}

\usepackage{graphicx}
\usepackage{epsfig}
\usepackage{epstopdf}

\newenvironment{Figure}
  {\par\medskip\noindent\minipage{\linewidth}\captionsetup{type=figure}}
  {\endminipage\par\medskip}

\newenvironment{Table}
  {\par\medskip\noindent\minipage{\linewidth}\captionsetup{type=table}}
  {\endminipage\par\medskip}

\title{\large \textbf{
Magnetic hopfions in solids
}}
\author[1,2]{\normalsize Filipp~N.~Rybakov\thanks{f.n.rybakov@gmail.com}}
\author[3]{\normalsize Nikolai~S.~Kiselev\thanks{n.kiselev@fz-juelich.de}}
\author[4,2]{\normalsize Aleksandr~B.~Borisov}
\author[5]{\normalsize Lukas D\"oring}
\author[5]{\normalsize\\Christof Melcher}
\author[3]{\normalsize Stefan Bl\"ugel}

\affil[1]{\small
Department of Physics, KTH-Royal Institute of Technology, SE-10691 Stockholm, Sweden}

\affil[2]{\small
Ural Federal University, Ekaterinburg 620002, Russia}

\affil[3]{\small
Peter Gr\"unberg Institut and  Institute for Advanced Simulation, Forschungszentrum J\"ulich and JARA, D-52425  J\"ulich, Germany}

\affil[4]{\small
M.N. Miheev Institute of Metal Physics of Ural Branch of Russian Academy of Sciences, Ekaterinburg 620990, Russia}

\affil[5]{\small
Department of Mathematics I \& JARA FIT, RWTH Aachen University, 52056 Aachen, Germany}

\date{}

\begin{document}

\maketitle
\vspace{-3.0\baselineskip}

\renewcommand{\abstractname}{\vspace{-\baselineskip}}
\begin{abstract}
Hopfions are an intriguing class of string-like solitons, named according to a classical topological concept classifying three-dimensional direction fields~\cite{Field_theory1}.
The search of hopfions in real physical systems is going on for nearly half a century, starting with the seminal work of Faddeev~\cite{Faddeev1975}. But so far realizations in solids are missing. 
Here, we present a  theory that identifies magnetic materials featuring hopfions as 
stable states without the assistance of confinement or external fields. 
Our results are based on an advanced micromagnetic energy functional derived from a spin-lattice Hamiltonian. 
Hopfions appear as emergent particles of the classical Heisenberg model. 
Magnetic hopfions represent three-dimensional particle-like objects of nanometre-size dimensions opening the gate to a new generation of spintronic devices in the framework of a truly three-dimensional architecture.
Our approach goes beyond the conventional phenomenological models. 
We derive material-realistic parameters that serve as concrete guidance in the search of magnetic hopfions   bridging computational physics with materials science. 
\end{abstract}

\renewcommand{\thesection}{\normalsize\Roman{section}}

\begin{multicols}{2}[]

\section{\normalsize Introduction}

Topological solitons~\cite{Field_theory1} are localized finite energy solutions to classical non-linear field equations appearing in many fields of science, from nuclear physics~\cite{Skyrme} to cosmic string theory~\cite{Vilenkin94}. Topological solitons are characterized by their ability to move and interact with each other as ordinary particles. 
They are one answer to Heisenberg's question how countable particles can appear in continuous fields. Recent celebrated examples in condensed matter are magnetic chiral skyrmions~\cite{Bogdanov_89, Yu_10, Melcher_14, Rybakov_15}, two-dimensional vortex-like field configurations in magnets without inversion symmetry, which have received considerable interest as fundamental objects as well as promising candidates for future spintronic applications~\cite{Fert_13}. 
In contrast, models, which allow three-dimensional (3D) topological solitons~\cite{Faddeev1975, Field_theory2}, to which we shall refer as hopfions according to a classical topological concept due to Heinz Hopf~\cite{Hopf31}, seem  very rare in nature. 
De facto, there is no evidence for stable hopfions in condensed matter physics appearing as freely movable objects. Instead, there are only a few experimental observations for objects that represent  Hopf fibrations in confined systems~\cite{Bouligand, Smalyukh2013, Hall16,  Ackerman_17}.

The occurence of stable hopfions in magnets has been predicted by Bogolubsky~\cite{Bogolubsky88} and Sutcliffe~\cite{Sutcliffe2017} based on a phenomenological micromagnetic model. Experimental evidence of magnetic hopfions, however, has never been reported, presumably because no relationship has yet been established between the parameters of the phenomenological models and the microscopic and atomistic nature of a solid. Here we introduce  a materials specific micromagnetic model and show that in magnetic materials characterized by competing magnetic exchange interactions, hopfions may appear as stable states and their stability does not require any assistance of confinement and external magnetic field. This advanced  micromagnetic model is derived from a microscopic spin-lattice Hamiltonian. The latter can  be deduced from the  quantum mechanical description of electrons in a solid typically with the help of density functional theory~\cite{Jones:15}. Such a computational materials science approach opens a multiscale vista for the discovery of hopfions by bridging quantum mechanics to higher-level materials specific models.

Topological solitons occur if the order parameter takes values in a curved manifold, e.g.\ a Lie group or a magnetization sphere $\mathbb{S}^2$. 
In this case, the topological concept of homotopy may be used to distinguish certain classes of field configurations from the topologically trivial uniform state. More precisely, for field configurations representing a topological soliton, there is no continuous deformation (homotopy) to any collinear (constant) field configuration.

In case of magnetic hopfions, the relevant order parameter of the system is a unit vector field $\mathbf{n}(\mathbf{r})=(n_x, n_y, n_z)$, $|\mathbf{n}(\mathbf{r})|=1$, defined at any point $\mathbf{r}\in\mathbb{R}^3$. Field configurations $\mathbf{n}$ attaining a uniform background state at infinity $\mathbf{n}(\mathbf{r}) \to \mathbf{n}_0$ 
as $|\mathbf{r}|\rightarrow\infty$ can be classified according to the linkage of their fibres $\{\mathbf{n} = \mathbf{p}\}$, which, for regular values $\mathbf{p} \in \mathbb{S}^2$, are collections of closed loops in $\mathbb{R}^3$.
Examples of simple and more intricate hopfion configurations are illustrated in Fig.~\ref{Fig_1}.

The Hopf invariant $H\!=\!H(\mathbf{n})$ of a field $\mathbf{n}$ onto $\mathbb{S}^2$ is defined to be the linking number of two generic fibres, i.e., the oriented number of times fibres wind around each other. The crucial point is that $H$ is a classifying homotopy invariant.
The existence of a homotopically non-trivial map with unit $H$, nowadays known as Hopf fibration, is a classical result due to Heinz Hopf~\cite{Hopf31} in 1931 also emerging implicitly in Dirac's work
on quantized singularities of the same year~\cite{Ryder:80}. 
Hopfions are topological solitons of non-zero Hopf invariant, whose existence or non-existence relies on the structure of the governing energy functional.

\begin{Figure}
\centering
\includegraphics[width=\columnwidth]{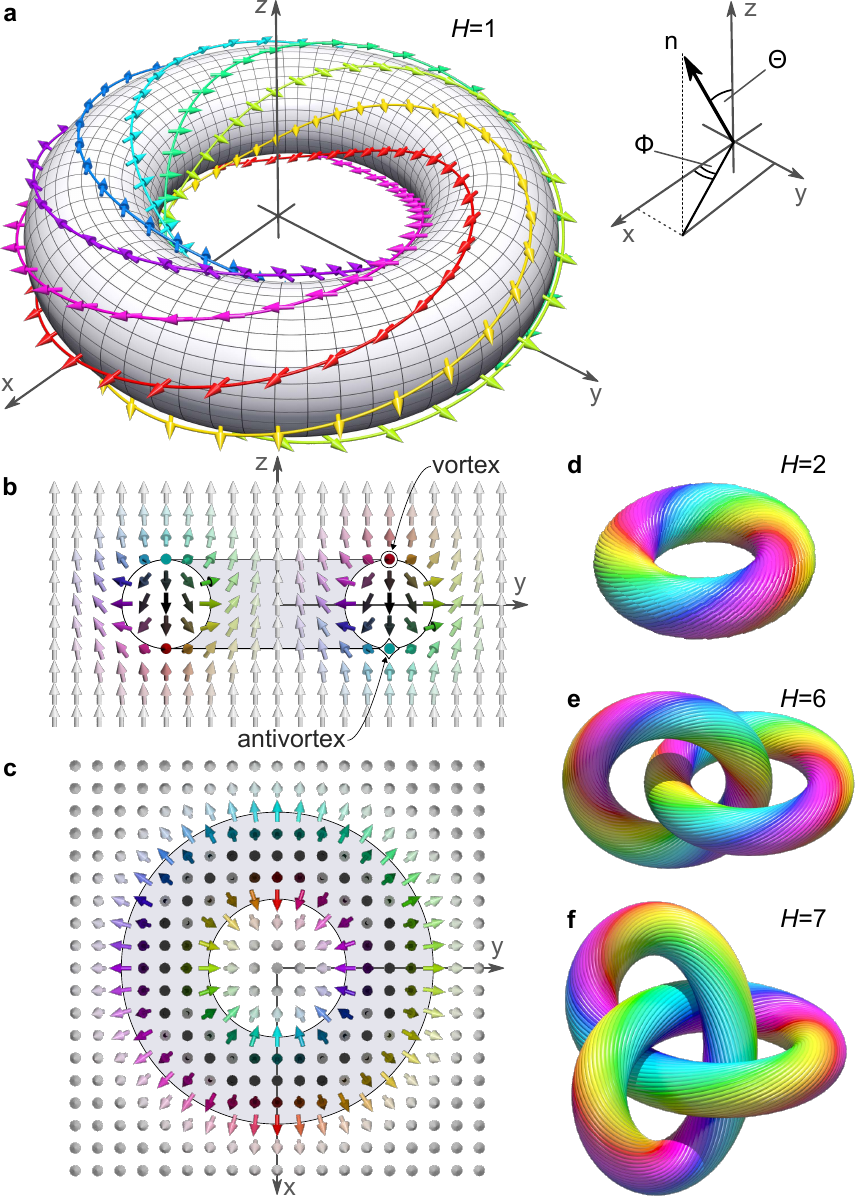}
\caption{\small \textbf{Typical structures of hopfions.}
\textbf{a}, Toroidal hopfion with Hopf index $H\!=\!1$. The solid colored isolines connect points $\mathbf{r}$ of a vector field $\mathbf{n}(\mathbf{r})=(\cos\Phi \sin\Theta,\sin\Phi\sin\Theta,\cos\Theta)$ with fixed values of angular variables, $\Theta\equiv\Theta(\mathbf{r})$ and $\Phi\equiv\Phi(\mathbf{r})$. The color of the vectors and the corresponding isolines are defined by the angular parameter $\Phi$ and together they compose the isosurface $\Theta\!=\!\textrm{const}$.
\textbf{b} and \textbf{c},  Texture of toroidal hopfion at intersecting planes $x\!=\!0$ and $z\!=\!0$, respectively.
\textbf{d}-\textbf{f}, Isosurfaces $\Theta\!=\!\textrm{const}$ for toroidal  hopfion with $H\!=\!2$, two linked toroidal-like hopfions with total $H\!=\!6$, and for hopfion with trefoil knot-like shape with $H\!=\!7$. 
}
\label{Fig_1}
\end{Figure}

It has generally been assumed that in magnetically ordered crystals such intricate 3D textures may only arise as precessing or moving states~\cite{Ivanov79, Papanicolaou93, Cooper99}, while energy dissipation processes finally lead to instability.
Comparing the energy density of the conventional micromagnetic theory~\cite{Aharoni}, $\mathcal{E}\!\propto\!(\mathbf{\nabla}\mathbf{n})^2\!\equiv\!\sum_{\alpha}(\nabla n_\alpha)^2$, $\alpha\!\in\!\{x, y, z\}$, with the original Skyrme model~\cite{Skyrme} one realizes that the soliton stabilizer term is missing in micromagnetic functionals, which require either higher powers of derivatives of the magnetization density, $\mathbf{n}(\mathbf{r})$, or higher order derivatives.

Here we show that magnetic hopfions can indeed exist as stable entities in magnetic materials with competing exchange interactions. We derive a continuum energy functional that is able to describe the competing interactions by virtue of additional stabilizer terms. The analysis of such an advanced micromagnetic functional provides simple criteria for the existence of hopfions.

\section{\normalsize Model}

The proposed functional arises from the classical isotropic Heisenberg model, a microscopic model defined through the following Hamiltonian:
\begin{equation}
\mathcal{H}=- \sum_{i>j}
\mathcal{J}_{ij} \, \mathbf{n}_i \! \cdot \! \mathbf{n}_j\, ,
\label{Ham1}
\end{equation}
which describes an exchange interaction between
magnetic spin moments $\boldsymbol{\mu}$ located on atomic sites $i$ and $j$, where $\mathbf{n}$ are unit vectors, $\mathbf{n}=\boldsymbol{\mu}/|\boldsymbol{\mu}|$. 
$\sum_{i>j}$ indicates the summation over all pairs of interacting spins.
Here we go beyond typical model approaches and consider the most general case of pair interactions  $\mathcal{J}_{ij}(\mathbf{r}_{ij})$ that do not vanish with distance as typical for metals~\cite{Pajda}.
The various potential energy terms such as magnetocrystalline anisotropy or Zeeman energy can be taken into account in further studies.
However, as shown below, they are not required for the stability of hopfions and thereby we have excluded them from the consideration.

\section{\normalsize Results and discussions}

We consider crystals with cubic Bravais lattices: simple (sc), body-centered (bcc) and face-centered (fcc).
For a correct continuum representation ($\mathbf{n}_i \rightarrow \mathbf{n}(\mathbf{r})$) of the spin-lattice Hamiltonian~\eqref{Ham1} which is able to reflect the essence of the competing exchange interactions, we go beyond the conventional micromagnetic approximation and take into account the higher order terms in the series expansion, see Section~\ref{methods}.
We obtain the advanced micromagnetic energy functional 
\begin{equation} 
E\!=\!\!\int\limits_{\mathbb{R}^3}
\!\mathcal{A} 
\Bigg(\! \frac{\partial \mathbf{n}}{\partial r_\alpha}  \!\Bigg)^{\!\!2}
\!\!+\mathcal{B} 
\Bigg(\!\frac{\partial^2 \mathbf{n}}{\partial r_\alpha^2} - \frac{\partial^2 \mathbf{n}}{\partial r_\beta^2} \! \Bigg)^{\!\!2}
\!\!+\mathcal{C}  
\Bigg(\! \frac{\partial^2 \mathbf{n}}{\partial r_\alpha \partial r_\beta} \!\Bigg)^{\!\!2}
d{\mathbf{r}}
\label{Func1}
\end{equation}
with implicit summations over indices $\alpha$ and $\beta\!\neq\!\alpha$, where each index runs over $x$, $y$, $z$. 
The derivation provides linear relations between micromagnetic and exchange constants: $\mathcal{A}\!=\!(1/a)\!\sum_{s}\! {\mathtt{a}_s J_s}$, $\mathcal{B}\!=\!-a\!\sum_{s}\! {\mathtt{b}_s J_s}$, $\mathcal{C}\!=\!-a\!\sum_{s}\! {\mathtt{c}_s J_s}$, 
where  $a$ is the lattice constant. The positive coefficients $\mathtt{a}_s$, $ \mathtt{b}_s$ and $\mathtt{c}_s$ depend on the crystal lattice type (see Tables~\ref{Tab_sc}, \ref{Tab_fcc}, \ref{Tab_bcc}).

For real materials the micromagnetic constants $\mathcal{A}$, $\mathcal{B}$, $\mathcal{C}$ can be  obtained either experimentally measuring spin-wave spectra along different crystallographic directions or using density functional theory (DFT) calculating the energy density for flat spin spiral textures~\cite{Heinze_11}, 
$\textbf{n}_\mathrm{s}=(\cos(\mathbf{q}\cdot\mathbf{r}),\sin(\mathbf{q}\cdot\mathbf{r}),0)$ with the wave vector $\mathbf{q}$, which reads:
\begin{align}
\mathcal{E}_\mathrm{s} = \mathcal{A}q^2 + \begin{cases}
\frac{2}{3}\mathcal{C}q^4 \quad 
&(\mathbf{q} \parallel [111])\, ,\\
4\mathcal{B}q^4 \quad 
&(\mathbf{q} \parallel [100])\, .
\end{cases}
\label{Qwave}
\end{align}
Fig.~\ref{Fig_min_mod} illustrates that, in contrast to conventional ferromagnets characterized by a quadratic law $\mathcal{E}_\mathrm{s}\sim q^2$, the functional~(\ref{Func1}) features quartic behaviour of $\mathcal{E}_\mathrm{s}$ near the $\Gamma$ point along at least one of the high symmetry directions. 
%

\begin{Figure}
\centering
\includegraphics[width=\columnwidth]{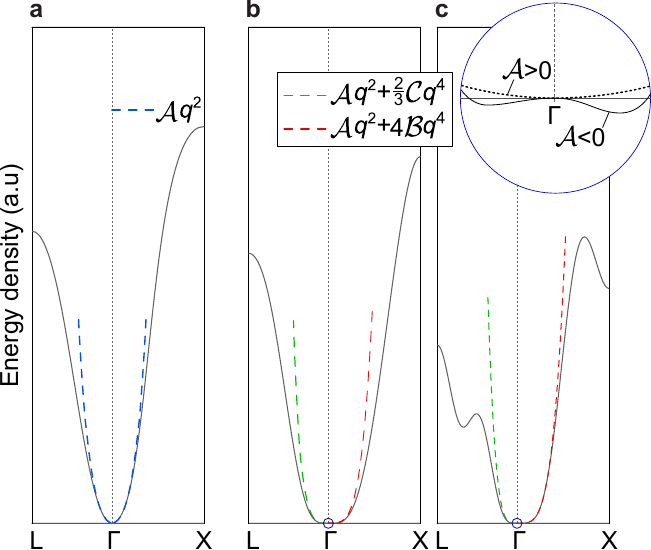}
\caption{\small \textbf{Examples of the dispersion curves for conventional magnets and for those that are able to host hopfions.} 
Examples of the energy density dependencies for the flat spin spirals with the wave vector along high symmetry directions of the cubic crystal for classical ferromagnet in \textbf{a} and for the systems which are able to host magnetic hopfions in
\textbf{b} and \textbf{c}. Dashed lines show the fit for the dispersion curves near the $\Gamma$ point. Inset in \textbf{c} shows possible behavior of $\mathcal{E}_\mathrm{s}$ shown in \textbf{b} and \textbf{c} in the close vicinity to the $\Gamma$ point. For $\mathcal{A}>0$ (dashed line) and for $\mathcal{A}<0$ (solid line) the global minimum corresponds to the collinear state ($q=0$) and spin spiral state ($q\neq0$), respectively. 
}
\label{Fig_min_mod}
\end{Figure}

For negative $\mathcal{B}$ or $\mathcal{C}$, the energy density~(\ref{Qwave}) is unbounded from below when $q\rightarrow\infty$, which indicates the tendency of the system to antiferromagnetic order where the functional~(\ref{Func1}) becomes irrelevant.
For positive $\mathcal{A}$, $\mathcal{B}$ and $\mathcal{C}$, the energy density in~(\ref{Func1}) is non-negative and provides a true extension of a conventional ferromagnet governed by $\mathcal{E}\!\propto\!(\mathbf{\nabla}\mathbf{n})^2$. 
In this case the functional~(\ref{Func1}) satisfies the Derrick-Hobart criterion~\cite{Field_theory1} necessary for the existence and stability of solutions localized in 3D space.

In the particular case $\mathcal{C}\!=\!6\mathcal{B}$ the energy density in \eqref{Func1} can be reduced to a simple form  $\mathcal{E}\!=\!\mathcal{A}(\mathbf{\nabla}\mathbf{n})^2\!+\!4\mathcal{B}(\Delta\mathbf{n})^2$ and becomes isotropic in real space. 
This canonical model has been studied before in Ref.~\cite{Bogolubsky88} 
and its particular case corresponding to negative $\mathcal{A}$ has been revisited in Ref.~\cite{Lin2016}.
The existence of hopfions in the presence of an additional potential energy term to suppress spin-spiral modulations has been reported recently in Ref.~\cite{Sutcliffe2017}.

From a mathematical perspective, the advanced micromagnetic model features striking analogies with the Skyrme-Faddeev model~\cite{Kapitanski79}. 
The latter one is accompanied by a well-established theory on the occurrence of string-like solitons. 
For positive coupling constants $\mathcal{A}$, $\mathcal{B}$, $\mathcal{C}$, we find a fractional power law for minimal energies $E\!\sim  |H|^\frac{3}{4}$ (see Fig.~\ref{Fig_E(H)} and Section~\ref{methods}). 
This is in contrast to the chiral skyrmion problem in 2D featuring linear energy growth with respect to the topological charge~\cite{Melcher_14}. 
By virtue of the arguments in Ref.~\cite{Lin_Yang:2004}, the sublinear energy growth essentially implies the attainment of minimal energies in infinitely many homotopy classes.

\begin{Figure}
\centering
\includegraphics[width=0.9\columnwidth]{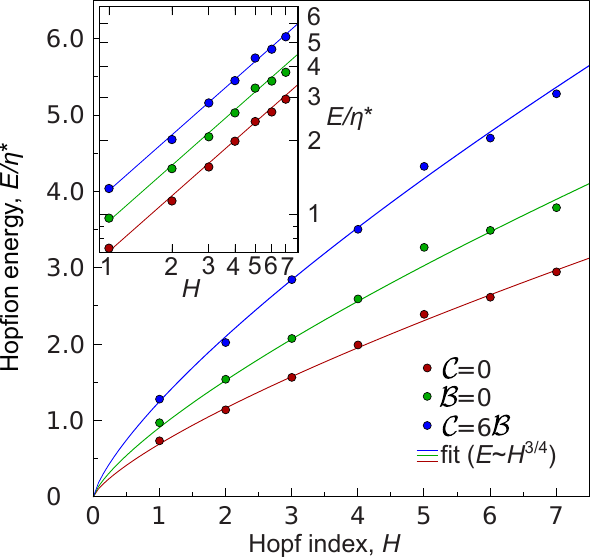}
\caption{\small \textbf{The energy of hopfions with different Hopf index. }
The energy of stable 3D solitons solutions as a function of the Hopf index. 
Hopfions have been numerically calculated in effective spin-lattice model with exchange coupling constants corresponding to three special cases: $\mathcal{B}=0$, $\mathcal{C}=0$ and $\mathcal{C}=6\mathcal{B}$. 
Inset shows the same dependencies given in logarithmic scale reproducing the power law: $E\sim |H|^{3/4}$.
Parameter  $\eta^{*}\!=\!389.3\!\cdot\!\sqrt{\mathcal{A}\cdot\max(\mathcal{C},6\mathcal{B})}$ is defined from the fit to the numerical calculations.
The morphology of solitons corresponding to each point is presented in Fig.~\ref{Fig_HopfZoo}. 
}
\label{Fig_E(H)}
\end{Figure}

The characteristic scale of inhomogeneity, $l_0$, which can be seen as a measure of smoothness of the hopfion's vector field, represents an important parameter related to the stability of such states.
The definition of $l_0$  and an analytical estimation in terms of the micromagnetic constants is given in Section~\ref{methods}.
Note, for strongly inhomogeneous textures, i.e.\ when $l_0\lesssim d/\delta_\mathrm{m}$, where $d$ is the nearest neighbor distance (sc: $d=a$, bcc: $d=a\sqrt{3}/2$, fcc: $d=a\sqrt{2}/2$) and $\delta_\mathrm{m}$ is the maximal angle between neighboring spins, which according to our estimation should not exceed approximately $45^\circ$ -- half of the critical angle for spin-lattice models~\cite{Ward06}, the continuum functional~(\ref{Func1}) may become an insufficient approximation of the lattice Hamiltonian~(\ref{Ham1}). 
The analytical expression of $l_0(\mathcal{A},\mathcal{B},\mathcal{C})$ together with above analysis of dispersion curves suggests the following criterion for the 
occurrence of hopfions
\begin{equation}
\max(\mathcal{C},6\mathcal{B}) 
\gtrapprox 
6.5 \mathcal{A}\,d^2, 
\label{cond}
\end{equation}
for $\mathcal{A}>0$ and $\mathcal{B},\mathcal{C}\ge 0$.
A remarkable property of~(\ref{Func1}) is that for a given set of material parameters $\mathcal{A}$, $\mathcal{B}$, $\mathcal{C}$ the numerical solutions can be found efficiently by means of an auxilliary \textit{minimal effective model} -- the spin lattice model for a simple cubic lattice with only four interacting shells of atoms (see Figure~\ref{Fig_S0}).
Indeed, for any system with sc, fcc or bcc lattice and a possibly infinite series of $\mathcal{J}_{ij}$, stable hopfion solutions can be calculated in this framework. 
In such a minimal model the effective coupling constants $\widetilde{J}_{1,2,3,4}$ are chosen in such a way that the ferromagnetic ground state and the corresponding macroscopic parameters $\mathcal{A}$, $\mathcal{B}$ and $\mathcal{C}$ remain the same.
This approach together with an advanced algorithm for  numerical energy minimization permits to identify a wide spectrum of solutions with different Hopf index.

\begin{figure*}
\centering
\includegraphics[width=16cm]{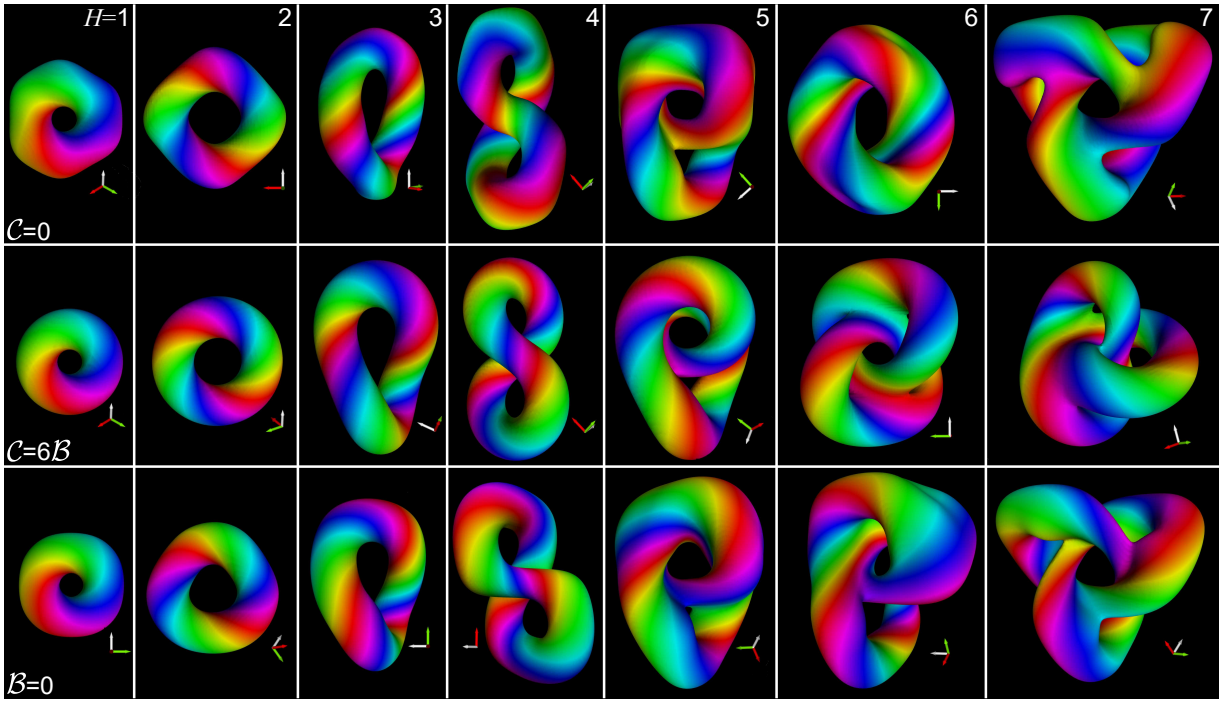}
\caption{\small \textbf{Morphology of magnetic hopfions.} Shown magnetization fields are energy minimizer of corresponding homotopy classes defined by the Hopf index $H$. Each row illustrates isosurfaces with $\Theta\!=\!\pi/2$ ($n_{z}\!=\!0$). Note, the complexity of the hopfion shape and size increases with the Hopf index. For $\mathcal{C}\!=\!6\mathcal{B}$ the hopfion with $H\!=\!6$ is similar to the linked tori state, while for
$\mathcal{C}\!=\!0$ and $\mathcal{B}\!=\!0$ the hopfion with $H\!=\!7$ are similar to trefoil knot, see Fig.~\ref{Fig_1}\,f. 
 }
\label{Fig_HopfZoo}
\end{figure*}

Figure~\ref{Fig_HopfZoo} illustrates hopfions for the limiting cases $\mathcal{C}=0$, $\mathcal{C}=6\mathcal{B}$, and $\mathcal{B}=0$. 
The axial symmetry of toroidal hopfions with Hopf index $H\!=\!1,\,2$  reflects the isotropic structure of functional \eqref{Func1} in case of $\mathcal{C}=6\mathcal{B}$.
Precisely this functional has been considered by Bogolubsky~\cite{Bogolubsky88} who found a stable hopfion with $H\!=\!1$ by means of numerical minimization of the energy of the corresponding effective spin-lattice model with $\widetilde{J}_1\!=\!61$, $\widetilde{J}_2\!=\!-10$, $\widetilde{J}_3\!=\!0$, $\widetilde{J}_4\!=\!-5$ given in arbitrary units, which correspond to the particular case of $\mathcal{C}\!\approx\!6\mathcal{B}$, see Figure~\ref{Fig_S0}.

Contrary to chiral magnetic skyrmions requiring materials with strong spin-orbit interaction in combination with a lattice lacking  inversion symmetry~\cite{Bogdanov_89,Yu_10,Heinze_11,Melcher_14}, the materials requirements for the existence of magnetic hopfions are much weaker. 
Thus, we expect experimental confirmation of their existence in the near future. 
%
Knowing criterion \eqref{cond}, motivates a DFT driven virtual design of  materials for stable hopfions \'a la carte~\cite{Dupe:16}.
X-ray magnetic nanotomography and off-axis electron holography are promising magnetic imaging techniques allowing reconstruction of 3D spin texture of hopfions.

Hopfions offer exciting perspectives in future information technology in which hopfions appear as the smallest 3D information carrying particles in spintronics. Electrons passing hopfions will experience an emergent magnetic field created by the  non-coplanar nature of the hopfion's magnetization texture giving rise to a hopfion Hall effect enabling the individual readout of Hopf invariants. Spin-currents can be used to move hopfions in all three directions leading eventually to the development of truly 3D memories. The non-linear dynamical response of  hopfions in a box offer fascinating perspectives for neuromorphic computing.

\section{\normalsize Methods}
\label{methods}

\noindent\textbf{Series expansion techniques.} 
We derive the continuum representation of the Heisenberg Hamiltonian~\eqref{Ham1} following the general micromagnetic approach~\cite{Aharoni}.
Since the coupling constants $\mathcal{J}_{ij}$ of the Heisenberg model~\eqref{Ham1} are symmetry related they can be decomposed into shells defined by the symmetry of the crystal lattice.
It means that for each $i$-th lattice site and its neighboring sites $k$ and $m$ the relation $\mathcal{J}_{ik} = \mathcal{J}_{im}$ is valid when the position vectors $\mathbf{r}_{ik}$ and $\mathbf{r}_{im}$ can be transformed into each other under the point group operations allowed by the symmetry of corresponding crystal. 
Complete group of symmetrically identical lattice sites compose a shell. Each of such shells is assigned to an integer index $s$ and corresponding coupling constant $J_s$. The shell of nearest neighbors has index  $s=1$, the shell of next nearest neighbors -- $s=2$ and so on.
Taking into account the fact that $\mathbf{n}$ represents a unit vector field, $\mathbf{n}^2=1$, the energy per one magnetic atom can be written as:
\begin{equation}
\epsilon\!=\!
\frac{1}{4}\!\sum_{s}
\!J_s\!\sum_{k,l,m}
\left[
\mathbf{n}\left(\mathbf{r}\!+\!a\,(k\mathbf{e}_x\!+\!l\mathbf{e}_y\!+\!m\mathbf{e}_z)\right) - \mathbf{n}(\mathbf{r})
\right]^2
,
\label{E_i}
\end{equation} 
where $\mathbf{r}$ is the position of the atom and $k,l,m$ are set of integer/half-integer indexes corresponding to the shell $s$ (see Tables~\ref{Tab_sc}, \ref{Tab_fcc}, \ref{Tab_bcc}).
The continuous form of the energy density can be obtained by expanding $\epsilon$ into a Taylor series with respect to $a$:
\begin{equation}
\mathcal{E} = \frac{1}{v} \sum_{k} \frac{a^k}{k!} \left( \frac{d^k \epsilon}{da^k} \right)_{a=0},
\label{series}
\end{equation}
where $v$ is the volume per single lattice site (sc: $v=a^3$, bcc: $v=a^3/2$, fcc: $v=a^3/4$).
It is easy to show that Eq.~(\ref{series}) with two leading terms leads to Eq.~(\ref{Func1}) (for details see Appendix~\ref{sup_theo_1}).
Note, only two leading terms in~(\ref{series}) which define both the structure and the scale of equilibrium magnetic textures are significant. 
Because the odd terms vanish due to the symmetry of the problem, we truncate the sum~(\ref{series}) at $k=5$.

\noindent\textbf{Calculation of the topological charge.}
The Hopf invariant of a unit vector field $\mathbf{n}(\mathbf{r})$ 
of appropriate smoothness and decay can be calculated
by means of the Whitehead formula~\cite{Whitehead47}: 
\begin{equation}
H=-\frac{1}{(8\pi)^2}\int_{\mathbb{R}^3}{(\mathbf{F}\cdot\mathbf{A})}\,\mathrm{d}\mathbf{r}.
\label{Hopf_index}
\end{equation}
The components of the solenoidal gyro-vector field $\mathbf{F}$ 
are explicitly determined by $\mathbf{n}$ and its derivatives, i.e.
$$F_i=\varepsilon_{ijk}
\mathbf{n}\cdot
\left[
\frac{\partial \mathbf{n}}{\partial r_j}
\times 
\frac{\partial \mathbf{n}}{\partial r_k}
\right]
,$$ 
where $\varepsilon_{ijk}$ is the 3D Levi-Civita symbol. The field $\mathbf{A}$ is an appropriate vector potential of $\mathbf{F}$, i.e., a decaying solution of
$$\nabla\times\mathbf{A}=\mathbf{F}$$
which may be computed in terms of elementary integrals, for example, like this
$$\mathbf{A}=\left(-\int\limits_{-\infty}^y\!F_z d{y},\,0,\,\int\limits_{-\infty}^y\!F_x d{y}\right).$$


\noindent\textbf{Topological energy bounds.}
By applying mathematical transformations (for details see Appendix~\ref{sup_theo_2}) we show that~\ref{Func1} is bounded from below by a variant of the Skyrme-Faddeev functional. 
The latter one has a topological lower bound, also known as Vakulenko-Kapitanski inequality~\cite{Kapitanski79}. 
After adapting constants~\cite{Rybakov81,Kundu_Rybakov}, we find 
\begin{equation}
E\geq\eta |H|^\frac34.
\nonumber
\end{equation}
with an explicit constant $\eta=(32\pi^2 / 3^{1/8})\sqrt{2\mathcal{A}\,\min(\mathcal{C},6\mathcal{B})}$, which is not expected to be optimal. 
This lower energy bound can be matched qualitatively by suitable trial configurations of prescribed Hopf invariant as in Ref.~\cite{Lin_Yang:2008}, proving the fractional energy law mentioned beforehand. 
During completion of this work, we became aware of the work~\cite{Harland} reporting similar results related to energy bounds, but for the case $\mathcal{A}\!<\!0$, which is not considered here.

\noindent\textbf{Evaluation of characteristic scale of inhomogeneity.}
The characteristic scale of inhomogeneity was estimated analytically by analyzing the structure of an infinite hopfion.
This technique is based on the approach proposed by L.D. Faddeev~\cite{Faddeev1975} where hopfion is considered as twisted and looped vortex tube~\cite{Enz,Vega}. 
For some $\sigma$-models~\cite{Rybakov81,Kundu_Rybakov}
this approach allows performing exact analytical analysis assuming that the length of such closed tube tends to infinity.
We adapted this technique for the case of spatially anisotropic Hamiltonian~(\ref{Func1}) and derived an analytical expression for estimate value of the characteristic scale of inhomogeneity
$$l_0 = \mathrm{min}
\left( 
\frac{|\mathbf{r}_2-\mathbf{r}_1|} {\angle(\mathbf{n}_2,\mathbf{n}_1)} 
\right) \approx 
0.5 \sqrt{ \max(\mathcal{C},6\mathcal{B})/\mathcal{A} }.
$$
where $\mathbf{n}_2$ and $\mathbf{n}_1$ are vectors at arbitrary positions $\mathbf{r}_2$ and $\mathbf{r}_1$, respectively and $\angle(\mathbf{n}_2,\mathbf{n}_1)$ is angle between these vectors.
The characteristic scale, $l_0 $ can be assumed to be nearly independent on Hopf index, $H$ while all hopfions  can be decomposed on similar structural elements, for details see Appendix~\ref{sup_theo_3}.

\noindent\textbf{Numerical energy minimization.}
For the energy minimization we use adaptive stereographic projections (for details see Supplementary Materials in~\cite{Rybakov_15}) and a nonlinear conjugate gradient method  which for the best performance has been parallelized to be used on GPU with NVIDIA CUDA architecture.
We performed full energy minimization starting with initial configuration suggested in~\cite{Fujii85}, which corresponds to the $H\!=\!1$ case, Fig.~\ref{Fig_1}\,a. In order to obtain hopfion with $H\!=\!2$ we place two instances of earlier minimized configuration for $H\!=\!1$ hopfion in the opposite corners of the large simulated domain of a cube shape and run the iterative minimization procedure allowing two hopfions to merge. In the following steps we follow the same procedure by merging hopfions with higher hopf indices. Finally, we got the full spectrum presented in Fig.~\ref{Fig_HopfZoo}.

The simulated domain was chosen to be a simple cubic lattice with a typical size of $256^3$ spins (lattice sites).
In the numerical energy minimization we have tested fixed and periodic boundary conditions and found that for a chosen size of simulated domain the solutions do not depend on the type of boundary conditions and on the further increase of the size of simulated domain. 
Thereby, found solutions is a true particlelike states which are able to be in a steady state or move and interact in three dimensions, contrary to imprintings of Hopf fibrations~\cite{Hall16, Ackerman_17, DMI_imprint1, DMI_imprint2, DMI_imprint3} and Hopf-like fibrations~\cite{Bouligand, Smalyukh2013} in confined systems.

For the calculations of hopfions in Fig.~\ref{Fig_HopfZoo} the following sets of parameters have been used; 
($\mathcal{B}\!=\!0$): $\widetilde{J}_1\!=\!1$, $\widetilde{J}_2\!=\!-0.25$, $\widetilde{J}_3\!=\!0.0004$, $\widetilde{J}_4\!=\!-0.0001$, 
($\mathcal{C}\!=\!6\mathcal{B}$): $\widetilde{J}_1\!=\!1$, $\widetilde{J}_2\!=\!0.2$, $\widetilde{J}_3\!=\!-0.273$, $\widetilde{J}_4\!=\!-0.174$,
($\mathcal{C}\!=\!0$): $\widetilde{J}_1\!=\!1$, $\widetilde{J}_2\!=\!0.3$, $\widetilde{J}_3\!=\!0.246$, $\widetilde{J}_4\!=\!-0.793$.
Note, from the wide range of allowable sets of effective coupling constants, see Figure~\ref{Fig_S0}, we have chosen some sets which may correspond to realistic exchange constants of well studied metallic magnets -- oscillatory in sign and near-exponentially decaying at finite temperature.

Our minimization technique breaks a high symmetry ansatz made for the initial guess and therefore we found states that are good candidates for minimizers over the corresponding homotopy classes.
Nevertheless, using numerical tools some uncertainties always remain. Similar uncertainty is also inherent to the numerical simulations of the well studied Skyrme-Faddeev model~\cite{Gladikowski, Sutcliffe07}, where few minimizers had been improved recently by means of alternative approaches~\cite{Jennings}.

\end{multicols}

\setcounter{section}{0}

\renewcommand{\thesection}{\normalsize\text{APPENDIX }\Alph{section}}

\renewcommand{\theequation}{\Alph{section}\arabic{equation}} 
\renewcommand{\thetable}{\Alph{section}\arabic{table}} 
\renewcommand{\thefigure}{\Alph{section}\arabic{figure}} 
\renewcommand{\thesubsection}{\normalsize\Alph{section}.\Roman{subsection}}

\part*{\centering \large Appendices}\label{app}

\section{\normalsize Relations between micromagnetic and atomistic constants}
\label{sup_tables}
\setcounter{equation}{0}
\setcounter{figure}{0}
\setcounter{table}{0}


\begin{Table}
\centering
\resizebox{\columnwidth}{!}{%
\begin{tabular}{|c|c|c|c|c|c|c|c|c|c|c|c|c|}
\hline
 & \multicolumn{12}{c|}{Shell index, $s$}\\
\cline{2-13}
 &1&2&3&4&5&6&7&8&9&10&11&12\\
\hline
$|k|,|l|,|m|$&{ 1, 0, 0}&{1, 1, 0}&{1, 1, 1}&{2, 0, 0}&{2, 1, 0}&{2, 1, 1}& {2, 2, 0}& {3, 0, 0} & {2, 2, 1} & {3, 1, 0}&{3, 1, 1}& {2, 2, 2} \\
\hline
$\sqrt{k^2\!+\!l^2\!+\!m^2}$&{1}&{$\sqrt{2}$}&{$\sqrt{3}$}&{2}&{$\sqrt{5}$}&{$\sqrt{6}$}&{$2\sqrt{2}$}&{3}&{3}&{$\sqrt{10}$}&{$\sqrt{11}$}&{$2\sqrt{3}$}\\
\hline
$\mathtt{a}_s$&{1/2}&{2}&{2}&{2}&{10}&{12}&{8}&{18}&{9/2}&{20}&{22}&{8}\\
\hline
$\mathtt{b}_s$&{1/96}&{1/24}&{1/24}&{1/6}&{17/24}&{3/4}&{2/3}&{11/8}&{27/32}&{41/12}&{83/24}&{2/3}\\
\hline
$\mathtt{c}_s$&{1/48}&{1/3}&{7/12}&{1/3}&{41/12}&{6}&{16/3}&{59/4}&{27/16}&{34/3}&{197/12}&{28/3}\\
\hline
\end{tabular}
}
\caption{\small\textbf{Relations between micromagnetic and atomistic constants for simple cubic (sc) lattice.} Coefficients given for the first 12 shells. 
\label{Tab_sc}}
\end{Table}


\begin{Table}
\centering
\resizebox{\columnwidth}{!}{%
\begin{tabular}{|c|c|c|c|c|c|c|c|c|c|c|c|c|}
\hline
 & \multicolumn{12}{c|}{Shell index, $s$}\\
\cline{2-13}
 &1&2&3&4&5&6&7&8&9&10&11&12\\
\hline
$|k|,|l|,|m|$&{1/2,\! 1/2,\! 0}&{1,\! 0,\! 0}&{1,\! 1/2,\! 1/2}&{1,\! 1,\! 0}&{3/2,\! 1/2,\! 0}&{1,\! 1,\! 1}& {3/2,\! 1,\! 1/2}& {2,\! 0,\! 0} & {3/2,\! 3/2,\! 0} & {2,\! 1/2,\! 1/2} & {2,\! 1,\! 0}&{3/2,\! 3/2,\! 1} \\
\hline
$\sqrt{k^2\!+\!l^2\!+\!m^2}$&{$1/\sqrt{2}$}&{1}&{$\sqrt{3/2}$}&{$\sqrt{2}$}&{$\sqrt{5/2}$}&{$\sqrt{3}$}&{$\sqrt{7/2}$}&{2}&{$3/\sqrt{2}$}&{$3/\sqrt{2}$}&{$\sqrt{5}$}&{$\sqrt{11/2}$}\\
\hline
$\mathtt{a}_s$&{2}&{2}&{12}&{8}&{20}&{8}&{56}&{8}&{18}&{36}&{40}&{44}\\
\hline
$\mathtt{b}_s$&{1/96}&{1/24}&{3/16}&{1/6}&{41/48}&{1/6}&{49/24}&{2/3}&{27/32}&{43/16}&{17/6}&{89/48}\\
\hline
$\mathtt{c}_s$&{1/12}&{1/12}&{3/2}&{4/3}&{17/6}&{7/3}&{49/3}&{4/3}&{27/4}&{19/2}&{41/3}&{137/6}\\
\hline
\end{tabular}
}
\caption{\small\textbf{Relations between micromagnetic and atomistic constants for face-centred (fcc) lattice.} Coefficients given for the first 12 shells. 
\label{Tab_fcc}}
\end{Table}


\begin{Table}
\centering
\resizebox{\columnwidth}{!}{%
\begin{tabular}{|c|c|c|c|c|c|c|c|c|c|c|c|c|}
\hline
 & \multicolumn{12}{c|}{Shell index, $s$}\\
\cline{2-13}
 &1&2&3&4&5&6&7&8&9&10&11&12\\
\hline
$|k|,|l|,|m|$&{1\!/2,\! 1\!/2,\! 1\!/2}&{1,\! 0,\! 0}&{1,\! 1,\! 0}&{3/2,\! 1\!/2,\! 1\!/2}&{1,\! 1,\! 1}&{2,\! 0,\! 0}& {3/2,\! 3/2,\! 1\!/2}& {2,\! 1,\! 0}& {2,\! 1,\! 1}   & {3/2,\! 3/2,\! 3/2} & {5/2,\! 1/2,\! 1/2} &{2,\! 2,\! 0} \\
\hline
$\sqrt{k^2\!+\!l^2\!+\!m^2}$&{$\sqrt{3}/2$}&{1}&{$\sqrt{2}$}&{$\sqrt{11}$/2}&{$\sqrt{3}$}&{2}&{$\sqrt{19}$/2}&{$\sqrt{5}$}&{$\sqrt{6}$}&{$3\sqrt{3}/2$}&{$3\sqrt{3}/2$}&{$2\sqrt{2}$}\\
\hline
$\mathtt{a}_s$&{1}&{1}&{4}&{11}&{4}&{4}&{19}&{20}&{24}&{9}&{27}&{16}\\
\hline
$\mathtt{b}_s$&{1/192}&{1/48}&{1/12}&{83/192}&{1/12}&{1/3}&{163/192}&{17/12}&{3/2}&{27/64}&{209/64}&{4/3}\\
\hline
$\mathtt{c}_s$&{7/96}&{1/24}&{2/3}&{197/96}&{7/6}&{2/3}&{757/96}&{41/6}&{12}&{189/32}&{311/32}&{32/3}\\
\hline
\end{tabular}
}
\caption{\small\textbf{Relations between micromagnetic and atomistic constants for body-centred (bcc) lattice.} Coefficients given for the first 12 shells. 
\label{Tab_bcc}}
\end{Table}


\section{\normalsize Minimal effective model}
\label{sup_toy_model}
\setcounter{equation}{0}
\setcounter{figure}{0}
\setcounter{table}{0}

\begin{Figure}
\centering
\includegraphics[width=12.65cm]{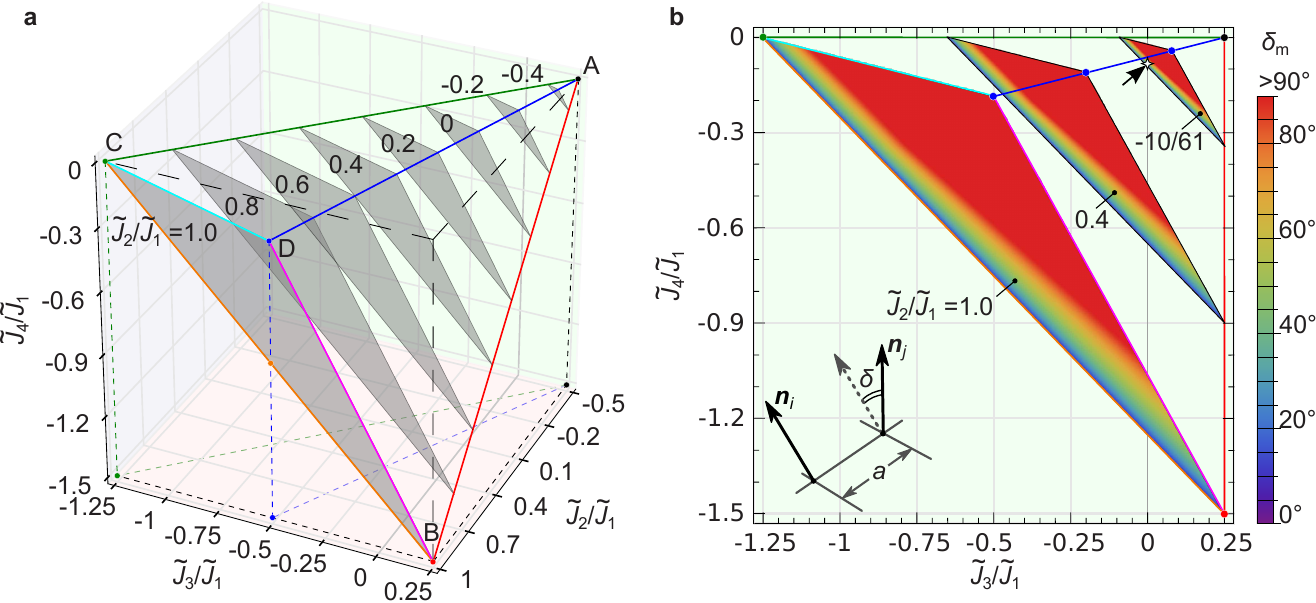}
\caption{\small \textbf{Range of existence of magnetic hopfions in minimal effective model.}
\textbf{a}, The micromagnetic parameters $\mathcal{A}$, $\mathcal{B}$ and $\mathcal{C}$ are positive when the reduced parameters for simple cubic lattice $\widetilde{J}_2/\widetilde{J}_1$, $\widetilde{J}_3/\widetilde{J}_1$, and $\widetilde{J}_4/\widetilde{J}_1$, ($\widetilde{J}_1>0$) belong to the domain restricted by three planes $ABC$, $ABD$ and $ADC$.
The shaded triangles represent the sections of the domain with fixed $\widetilde{J}_2/\widetilde{J}_1$.
\textbf{b}, Alternative representation of the diagram shown in \textbf{a}  obtained by projection into the plane of $\widetilde{J}_3/\widetilde{J}_1$ and $\widetilde{J}_4/\widetilde{J}_1$. 
As in \textbf{a}, the triangles corresponds to fixed values of $\widetilde{J}_2/\widetilde{J}_1$.
The color map indicates an analytical estimate for the maximal angle $\delta_\mathrm{m}$ between nearest $\textbf{n}$ vectors, see inset. 
The small arrow for $\widetilde{J}_3/\widetilde{J}_1$=0 indicates the single point found by Bogolubsky~\cite{Bogolubsky88}, for nearly isotropic case. 
}
\label{Fig_S0}
\end{Figure}

\section{\normalsize Theoretical analysis}
\setcounter{equation}{0}
\setcounter{figure}{0}
\setcounter{table}{0}

\subsection{\normalsize Continuous formula for the energy functional}
\label{sup_theo_1}

Summation of the two leading terms in the equation~(\ref{series}) of the Section~\ref{methods} gives 
\begin{equation}
\mathcal{E} = \mathcal{A}\,\mathcal{E}_A + \mathcal{B}\,\mathcal{E}_B + \mathcal{C}\,\mathcal{E}_C + 
\mathcal{A}_0\,\mathcal{E}_{A_0} + \mathcal{B}_0\,\mathcal{E}_{B_0} + \mathcal{C}_0\mathcal{E}_{C_0},
\label{fullE}
\end{equation}
where
\begin{eqnarray}
\mathcal{E}_A &=& \sum_{\alpha,\beta}
\left(\frac{\partial \, n_\alpha}{\partial r_\beta}  \right)^2,\\
\mathcal{E}_B &=& \sum_{\alpha,\beta\neq\gamma}
\left(\frac{\partial^2 \, n_\alpha}{\partial r_\beta^2}
-\frac{\partial^2 \, n_\alpha}{\partial r_\gamma^2}  \right)^2,\\
\mathcal{E}_C &=& \sum_{\alpha,\beta\neq\gamma}
\left(\frac{\partial^2 n_\alpha}{\partial r_\beta \partial r_\gamma} \right)^2,\\
\mathcal{E}_{A_0} &=& \sum_{\alpha,\beta}
\frac{\partial^2 }{\partial r_\beta^2}
\left(\frac{\partial \, n_\alpha}{\partial r_\beta}  \right)^2,\\
\mathcal{E}_{B_0} &=& \sum_{\alpha,\beta\neq\gamma}
\frac{\partial^2 }{ \partial r_\beta \partial r_\gamma}
\left(
\frac{\partial \, n_\alpha}{\partial r_\beta}  
\cdot
\frac{\partial \, n_\alpha}{\partial r_\gamma} 
\right),\\
\mathcal{E}_{C_0} &=& \sum_{\alpha,\beta\neq\gamma}
\frac{\partial^2 }{\partial r_\beta^2}
\left(\frac{\partial \, n_\alpha}{\partial r_\gamma}  \right)^2,\\
\mathcal{A}_0 &=& -8\mathcal{B}, \quad \mathcal{B}_0=4\mathcal{B}-\mathcal{C}, \quad \mathcal{C}_0=-\mathcal{C},
\end{eqnarray}
where $\mathcal{A}$, $\mathcal{B}$ and $\mathcal{C}$ are the linear functions on coupling constant $J_s$.
After the integration of~(\ref{fullE}) the last three terms $\mathcal{E}_{A_0}$, $\mathcal{E}_{B_0}$ and $\mathcal{E}_{C_0}$ can be ignored for bulk solids, because corresponding expressions represent total derivatives that can be reduced the integrals over the sample surface, which are only relevant for small enough specimens. 
Finally, the energy functional is
\begin{equation}
E=\int_{\mathbb{R}^3}\left(\mathcal{A}\,\mathcal{E}_A+\mathcal{B}\,\mathcal{E}_B+\mathcal{C}\,\mathcal{E}_C \right)\mathrm{d}\mathbf{r}.
\label{E}
\end{equation}


\subsection{\normalsize Bounding from below by a variant of the Skyrme-Faddeev functional}
\label{sup_theo_2}

For the unit vector fields we have inequality~\cite{Kittel49}:
\begin{eqnarray}
\mathcal{E}_A=-\mathbf{n}\cdot\Delta\mathbf{n},
\label{EA_Cartesian}
\end{eqnarray}
where $\Delta$ 
is the Laplace operator.
Moreover, in terms of angular variables $\mathcal{E}_A$ can be written as:
\begin{eqnarray}
\mathcal{E}_A=\left( |\nabla\Theta| - |\sin(\Theta)\nabla\Phi| \right)^2 + 2 |\nabla\cos(\Theta)| |\nabla\Phi|,
\label{EA_spherical}
\end{eqnarray}
where $\Theta\equiv\Theta(\mathbf{r})$ and $\Phi\equiv\Phi(\mathbf{r})$ angular variables defining vector field $\mathbf{n}=(\cos\Phi\, \sin\Theta,\sin\Phi\,\sin\Theta,\cos\Theta)$.
The gyro-vector field $\mathbf{F}$ which components are defined as
$$F_i=\sum_{j,k} \varepsilon_{ijk}
\mathbf{n}\cdot
\left[
\frac{\partial \mathbf{n}}{\partial r_j}
\times 
\frac{\partial \mathbf{n}}{\partial r_k}
\right]
$$ 
can also be written in angular variables:
\begin{eqnarray}
\mathbf{F}= -2 [ \nabla\cos(\Theta) \times \nabla\Phi ].
\label{F_spherical}
\end{eqnarray} 
Comparing the expression for $\mathcal{E}_A$ in  \eqref{EA_spherical} with the modulus of the vector field $\mathbf{F}$ in \eqref{F_spherical} (or in other way~\cite{Ward:1999}) one can show that the following inequality  is satisfied
\begin{eqnarray}
|\mathbf{F}|\leq\mathcal{E}_A 
\nonumber
\end{eqnarray} 
and taking into account \eqref{EA_Cartesian} we also obtain:
\begin{eqnarray}
|\mathbf{F}|\leq\mathcal|\Delta\mathbf{n}| 
\label{F<dn}
\end{eqnarray} 

One can show that the energy density terms with the higher order derivatives in~(\ref{E}) can be written in the following form:
\begin{eqnarray}
\mathcal{B}\,\mathcal{E}_B + \mathcal{C}\,\mathcal{E}_C
&=&\frac{2}{3}\min(\mathcal{C},6\mathcal{B})\, (\Delta{\mathbf{n}})^2 \nonumber\\ 
&+&\frac{1}{6}\max(0,6\mathcal{B}-\mathcal{C})\, \mathcal{E}_B +
\max(0,\mathcal{C}-6\mathcal{B})\, \mathcal{E}_C \nonumber\\
&+&\min(\mathcal{C},6\mathcal{B})\, \left( \Delta{\mathcal{E}_A} - \mathcal{E}_D  \right), 
\label{EB+EC}
\end{eqnarray} 
where 
\begin{equation}
\mathcal{E}_D = \sum_{\alpha,\beta,\gamma}
\frac{\partial^2 }{ \partial r_\beta \partial r_\gamma} \left(  \frac{\partial \, n_\alpha}{\partial r_\beta} \frac{\partial \, n_\alpha}{\partial r_\gamma} \right).
\nonumber
\end{equation}
The equality~\eqref{EB+EC} can be easily confirmed for both cases, $\mathcal{C}<6\mathcal{B}$ and $\mathcal{C}\geq6\mathcal{B}$. 
The last term in \eqref{EB+EC} is a boundary term and vanishes 
after the integration over the whole space $\mathbb{R}^3$ since  $\mathbf{n}|_{|\mathbf{r}|\rightarrow\infty}=\mathbf{n}_0$. The remaining terms in \eqref{EB+EC} are non-negative.
Taking into account \eqref{F<dn} one deduces that the functional~(\ref{E}) is bounded from below as follows:
\begin{equation}
\phantom{\leq}\int_{\mathbb{R}^3}\big( \mathcal{A}\,\mathcal{E}_A + \frac{2}{3}\min(\mathcal{C},6\mathcal{B}\big) \mathbf{F}^2) \,\mathrm{d}\mathbf{r}
\leq
\int_{\mathbb{R}^3}( \mathcal{A}\,\mathcal{E}_A + \mathcal{B}\,\mathcal{E}_B + \mathcal{C}\,\mathcal{E}_C ) \,\mathrm{d}\mathbf{r}\, .
\label{lowerbound}
\end{equation}
The left-hand side of this inequality coincides with a variant of the Skyrme-Faddeev functional.

\subsection{\normalsize Evaluation of characteristic scale of inhomogeneity for hopfion textures}
\label{sup_theo_3}

\begin{figure*}[ht]
\centering
\includegraphics[width=14cm]{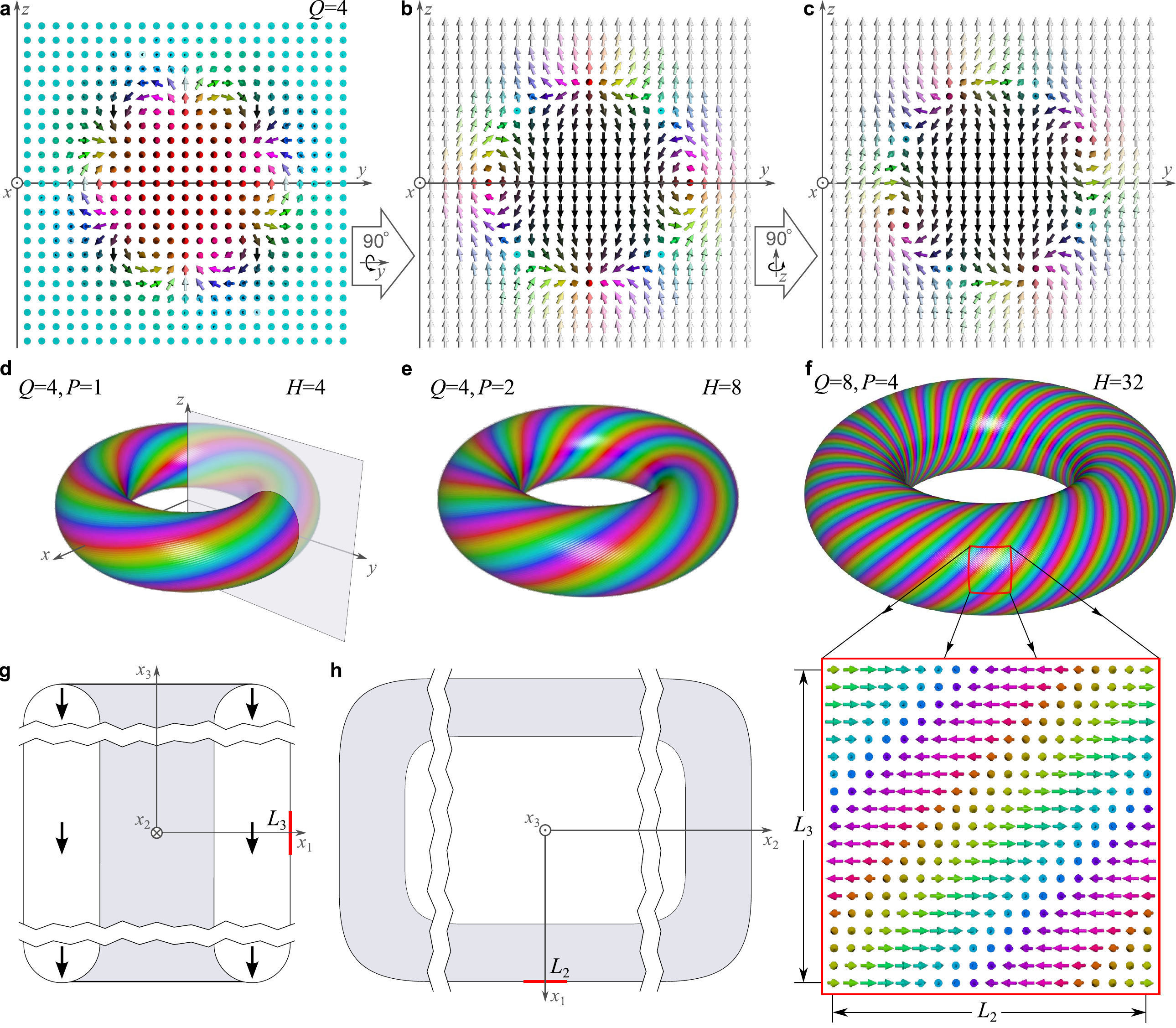}
\caption{\small \textbf{Construction of infinite hopfion for spatially anisotropic Hamiltonian.}
\textbf{a}, Spin texture of ordinary 2D vortex with topological charge $Q=4$ defined in $zOy$-plain.
\textbf{b}, Oblique 2D vortex obtained from ordinary vortex in \textbf{a} by means of homogeneous rotation of each spin around its own axis parallel to the $y$-axis.
\textbf{c}, Oblique 2D vortex with additional twist of each spin by $\pi/2$ around the own axis parallel to its asymptotic, $\mathbf{n}_0 \parallel \mathbf{e_z}$. 
\textbf{d}-\textbf{f}, Iso-surfaces of toroidal axisymmetric hopfions, $\Theta=\pi/2$.
\textbf{d}, Hopfion with $H=P \cdot Q =\!4$ constructed by rotation of oblique vortex in \textbf{b} around the  \textit{z}-axis. 
\textbf{e}, Hopfion with $H=\!8$ constructed by rotation of vortex in \textbf{b} around the  \textit{z}-axis and additional rotation of each spin by $2\pi$ ($P=2$) around $\mathbf{n}_0$.
\textbf{f}, Hopfion with $H\!=\!32$, constructed similar to that in \textbf{e}.
Inset shows unit cell of the domain wall, with periodicity $L_2$ and $L_3$ in two directions. 
\textbf{g} and \textbf{h}, Schematic representation of the shape (iso-surface) of infinite hopfion. The energy contribution of curved part of hopfion can be neglected when indecies $Q$ and $P\rightarrow\infty$.
}
\label{Fig_S1}
\end{figure*}

Let us consider an ordinary $\pi$-vortex spin texture~\cite{Piette} defined in a half-plane (Fig.~\ref{Fig_S1}\,a).
Oblique vortex with asymptotic $\mathbf{n}_0=(0,0,1)$ ($\Theta=0$) for $y\!=\!0$, $y\!\rightarrow\!\infty$, $z\!\rightarrow\!\pm\infty$ in (Fig.~\ref{Fig_S1}\,b) can be obtain via the rotation of each spin in (Fig.~\ref{Fig_S1}\,a) around own axis parallel to $\mathbf{e}_y$. 
The spin texture of such oblique 2D vortex can be considered as a bound state of half lumps pairs~\cite{JS_2010, Garaud2012} (vortex-antivortex pairs). 
Note, under such transformation the topological charge of oblique 2D vortex remains equal to the topological charge of initial vortex.
The hopfion texture in (Fig.~\ref{Fig_S1}\,d) can be constructed via the rotation of ``frozen'' texture in (Fig.~\ref{Fig_S1}\,b) around the $z$-axis. In this case, because of the global rotation, each spin experience $2\pi$-twisting (rotation) around their own axis parallel to $\mathbf{n}_0$. 
Usually such a twisting is defined by integer index $P$. 
For hopfion in (Fig.~\ref{Fig_S1}d) it equals unity meaning that each individual spin of the vortex texture is rotated by $2\pi$ only. 
Hopfions with higher topological index (Fig.~\ref{Fig_S1}\,e, f) can be constructed by $2\pi\!\cdot\!P$ twisting, $P\!>\!1$.
In frame of SF model particular cases where both  $P$ and $Q$ are small integer numbers have been studied numerically~\cite{Hietarinta, Kobayashi2014}. 
Earlier, it had been shown that for the functionals based on SF model the construction of axisymmetric toroidal hopfions with $Q=1$, $P\rightarrow\infty$ allows an analytic analysis~\cite{Kundu_Rybakov_sup}. 
Here, however, we are not able to apply the same approach because of the natural anisotropy of our model Hamiltonian which in most general case does not allow the existence of axisymmetric solutions.
Therefore, we introduce a new approach base on construction of hopfion with both $Q$ and $P\rightarrow\infty$ and shape schematically depicted in (Fig.~\ref{Fig_S1}\,g, h).

Let us consider a Cartesian coordinate system $(x_1,x_2,x_3)$ which orientation relative to crystallographic directions is described by the rotation matrix $R$:
\begin{eqnarray}
\begin{pmatrix}
 x_1 \\
 x_2 \\
 x_3
\end{pmatrix}
=
R \cdot 
\begin{pmatrix}
 x \\
 y \\
 z
\end{pmatrix},
\end{eqnarray}
where $\mathbf{x}\!\parallel\![100]$, $\mathbf{y}\!\parallel\![010]$, $\mathbf{z}\!\parallel\![001]$.

For certain orientations, the following kind of domain wall can be found as a solution of corresponding Euler-Lagrange equations for the functional~(\ref{E}):
\begin{eqnarray}
\Theta = \Theta(x_1), \quad \Phi = \frac{2\pi}{L_{2}}x_2 + \frac{2\pi}{L_{3}}x_3,
\label{FlatSolution}
\end{eqnarray}
where $L_{2}$ and $L_{3}$ are arbitrary parameters defining the period of modulations along each direction and $\Theta(x_1)$ - kink-type solution of corresponding ordinary differential equation: 
\begin{equation*}
\Theta(x_1) = 
 \begin{cases}
  \pi & (x_1\rightarrow-\infty)\\
  0 & (x_1\rightarrow\infty)
 \end{cases}
\end{equation*}

For $L_2\rightarrow\infty$  such solution describes the straight domain wall with periodic kinks~ \cite{Kobayashi2013} and can be considered as an element of 2D vortex~\cite{Piette, Garaud2012} with topological index $Q\rightarrow\infty$ and negligible curvature of the wall splitting domains with $\Theta\!=\!\pi$ and $\Theta\!=\!0$. 
When $L_2$ is finite (Fig.~\ref{Fig_S1}\,f), the solution \eqref{FlatSolution} describes an inhomogeneous tube which cross-section represents not circular immense 2D vortex while $\mathbf{n}$-vectors additionally twisted by $2\pi$ around own axes parallel to $\mathbf{n}_0$ for any $\Delta{x_2}\!=\!L_2$ shift. 
On the other hand, it can be considered as closed tube of small curvature -- \textit{infinite hopfion} with $H=P\cdot Q$, where $P,Q\rightarrow\infty$ (Fig.~\ref{Fig_S1}\,g, h).

Because the solution for $\Theta(x_1)$ is exponentially localized one can consider the element of such texture bounded by $0\!\leqslant\!x_2<\!L_2$ and $0\!\leqslant\!x_3<\!L_3$ as unit cell of which the whole infinite hopfion is composed.
In the limiting case for such infinite hopfion the contribution of bended domains (top and bottom sections in Fig.~\ref{Fig_S1}\,g, left and right -- in Fig.~\ref{Fig_S1}\,h) can be neglected. 
Similar to the Derrick-Hobart approach applied for the whole hopfion texture one can introduce similar criteria for the spin texture in such unit cell too:
\begin{eqnarray}
\left( \frac{d}{d\lambda}\int \mathcal{E}\left\{ \mathbf{n}(\lambda x_1,\lambda x_2,\lambda x_3) \right\} d\mathbf{r} \right)_{\lambda=1} &= 0 \nonumber\\
\left( \frac{d}{d\lambda}\int \mathcal{E}\left\{ \mathbf{n}(x_1, x_2,\lambda x_3) \right\} d\mathbf{r} \right)_{\lambda=1} &= 0 \label{HDarguments}
\end{eqnarray}

Because the maximal rotation of $\mathbf{n}$ for infinitesimal distance is reached for $x_1=0$, the characteristic scale of inhomogeneity
\begin{equation}
l_0 = \left(  {\Theta^\prime(0)}^2 + 4\pi^2\left( \frac{1}{{L_2}^2} + \frac{1}{{L_3}^2} \right)  \right)^{-1/2},
\label{DeltaEstimation}
\end{equation}
where $L_2$ and $L_3$ are defined by the system of equations \eqref{HDarguments}. Because all hopfions has the similar structure of the domain wall with slightly different distortions, which nevertheless do not affect the $l_0$ value sufficiently, the expression~(\ref{DeltaEstimation}) derived for the infinite hopfion can be used as good estimation for any hopfion solution with $H\in\mathbb{Z}_{\neq 0}$.

The integration of corresponding fourth order nonlinear ordinary differential equation for $\Theta(x_1)$ represents serious issue. Because of that we have fitted solution by the following analytical expression:
\begin{equation}
\Theta(x_1) \approx \arccos \left( \tanh \left( \frac{x_1}{\sqrt{\xi}} \right) \right).
\end{equation}

\textbf{Case $\mathcal{C} \leqslant 6\mathcal{B}$.}

The rotation matrix 
\begin{eqnarray}
R = \frac{1}{2}
\begin{pmatrix}
 \sqrt{2} & 1 & 1 \\
 0 & \sqrt{2} & -\sqrt{2} \\
 -\sqrt{2} & 1 & 1
\end{pmatrix}.
\end{eqnarray}

We got $L_2 = L_3$. The corresponding estimation for $l_0$ is 
\begin{eqnarray}
l_0 \approx \sqrt{\frac{\mathcal{B}}{\mathcal{A}}} \sqrt{
\frac
{x\,(-14+6c+3x)}
{-8+24c+3x}
},
\end{eqnarray}
where $c=\mathcal{C}/(6\mathcal{B})$ and $x$ is the real positive root of the third degree polynomial equation:
\begin{equation}
-160 + 864 c^3 - 144 c^2 (-10 + x) - 112 x + 75 x^2 - 9 x^3 + c (-96 + 384 x - 63 x^2) = 0.
\end{equation}
For all admissible $c$, $0\leqslant c \leqslant 1$, the value of $l_0$ lies in a relatively narrow range:
\begin{equation}
1.24 \lessapprox l_0\sqrt{\frac{\mathcal{A}}{\mathcal{B}}}  \lessapprox 1.66. \label{est1}
\end{equation}

\textbf{Case $6\mathcal{B} \leqslant \mathcal{C}$.}

The rotation matrix is trivial
\begin{eqnarray}
R = 
\begin{pmatrix}
 1 & 0 & 0 \\
 0 & 1 & 0 \\
 0 & 0 & 1
\end{pmatrix}.
\end{eqnarray}

We got $L_2 = L_3$. The corresponding estimation for $l_0$ is 
\begin{eqnarray}
l_0 \approx \sqrt{\frac{\mathcal{C}}{\mathcal{A}}} \sqrt{
\frac
{x\,(-6 + 2 b + 9 x)}
{-6 + 14 b + 9 x}
},
\end{eqnarray}
where $b=(6\mathcal{B})/\mathcal{C}$ and $x$ is the real positive root of the third degree polynomial equation:
\begin{eqnarray}
64 b^3 + b^2 (264 - 84 x) - 27 (2 - 3 x)^2 x -  18 b (4 - 16 x + 15 x^2) = 0.
\end{eqnarray}
For all admissible $b$, $0\leqslant b \leqslant 1$, the value of $l_0$ lies in a relatively narrow range:
\begin{equation}
0.68 \lessapprox l_0\sqrt{\frac{\mathcal{A}}{\mathcal{C}}}  \lessapprox 0.82. \label{est2}
\end{equation}

Combining~(\ref{est1}) and~(\ref{est2}) we can estimate the minimum value of $l_0$:
\begin{equation}
l_0 \approx 0.5 \sqrt{ \max(\mathcal{C},6\mathcal{B})/\mathcal{A} }.
\end{equation}

\part*{\centering \large Acknowledgements}

We thank O.~Chugreeva, B.~Dup\'e, P.F.~Bessarab, J.~Garaud, A.~Samoilenka, E.~Babaev  for discussions.
Ch.M.\ acknowledges funding from Deutsche Forschungsgemeinschaft (DFG grant no.\ ME 2273/3-1).
S.B.\ acknowledges funding from DFG through SPP 2137 ``Skyrmionics" (grant no.\  BL 444/16-1), the Collaborative Research Center SFB 1238 (Project C01) and through the DARPA TEE program through grant MIPR\# HR0011831554 from DOI.
L.D., Ch.M.\ and S.B.\ acknowledge seed-fund support from JARA-FIT.
The work of F.N.R. was supported by the Swedish Research Council Grant No. 642-2013-7837 and by G\"{o}ran Gustafsson Foundation for Research in Natural Sciences and Medicine. 
The  research  of A.B.B. was carried out within the state assignment of the Ministry of Education and Science (the theme ``Quantum'', number АААА-А18-118020190095-4).

\part*{\centering \large References}

\renewcommand{\section}[2]{}

\end{document}